\documentclass[10pt,letterpaper]{article}

\usepackage[margin=1in]{geometry}
\usepackage[utf8]{inputenc}
\usepackage[T1]{fontenc}
\usepackage{graphicx}
\usepackage{amsmath,amssymb,amsthm}
\usepackage{booktabs}
\usepackage{algorithmicx}
\usepackage{algpseudocode}
\usepackage{subcaption}
\usepackage{csquotes}
\usepackage{comment}
\usepackage[numbers,sort&compress]{natbib}
\usepackage{xcolor}
\usepackage{hyperref}
\usepackage{tcolorbox}
\tcbuselibrary{skins,breakable}
\newenvironment{keywords}{\par\smallskip\noindent\textbf{Keywords: }\ignorespaces}{\par\smallskip}

\begin{document}

\title{Collective Opinion Dynamics in Structured LLM Populations}
\date{}
\author{%
Erica Cau$^{1,2}$\thanks{Affiliations: 1 Department of Computer Science, University of Pisa; 2 ISTI-CNR. ORCID: 0009-0005-7267-9870. Email: \texttt{erica.cau@phd.unipi.it}.}%
\and Andrea Failla$^{2}$\thanks{Corresponding author. Affiliations: 2 ISTI-CNR. ORCID: 0009-0009-6162-0274. Email: \texttt{andrea.failla@isti.cnr.it}. Website: \url{https://andreafailla.github.io}.}%
\and Giulio Rossetti$^{2}$\thanks{Affiliation: 2 ISTI-CNR. ORCID: 0000-0003-3373-1240. Email: \texttt{giulio.rossetti@isti.cnr.it}. Website: \url{https://giuliorossetti.net}.}%
}
\maketitle

\begin{abstract}
\noindent Large Language Models are increasingly deployed as interacting agents in settings such as online platforms, recommendation systems, and multi-agent applications. 
Understanding the collective behaviors that emerge from their interactions is therefore increasingly crucial, especially as these behaviors may shape public opinion and contribute to polarization.
In this work, we investigate how network structure and group composition shape the evolution of opinions in populations of LLM agents engaged in multi-round debates.
We generate five networks with controlled levels of homophily and varying group sizes, and perform ten independent runs per configuration.
The results show that LLM agents exhibit patterns that are highly sensitive to network structure, relative group sizes, and to the model itself. 
We highlight that even within LLM populations, low homophily facilitates convergence by increasing opportunities for cross-group interaction, whereas higher homophily limits cross-group exposure and can preserve distinct opinion states, consistent with previous research.
We further show that the choice of LLM substantially affects the resulting dynamics, with different models exhibiting different patterns of opinion updating under comparable network conditions.
At the individual level, providing agents with information about their local neighborhood further modulates opinion transitions, revealing substantial differences between models in their sensitivity to local social context.
Overall, our findings show that the collective dynamics of LLM populations arise from interactions among model-specific behavior, network structure, population composition, and local social information.
\end{abstract}

%\begin{graphicalabstract}
%\includegraphics{figs/cas-grabs.pdf}
%\end{graphicalabstract}

%\begin{highlights}
%\item Research highlights item 1
%\item Research highlights item 2
%\item Research highlights item 3
%\end{highlights}

% Use if graphical abstract is present
%\begin{graphicalabstract}
%\includegraphics{}
%\end{graphicalabstract}

% Research highlights
%\begin{highlights}
%\item 
%\item 
%\item 
%\end{highlights}

%\nocite{*}

% Keywords
% Each keyword is seperated by \sep
%\linenumbers
% Main text
\section{Introduction}\label{sec:intro}
%\paragraph{LLM intro}
The introduction of Large Language Models (LLMs)~\cite{singh2025openai, jiang2023mistral7b,llama3modelcard,deepseekai2025deepseekv3technicalreport} has revolutionized the field of Natural Language Processing (NLP), advancing the ability of language models to generate text that closely resembles human writing~\cite{clark2021all}. 
{Through these skills, the models have demonstrated unprecedented capabilities, even in fields not closely related to linguistics, such as medicine \cite{wang2026human, wang2025survey}, law \cite{liu2026llm}, code generation \cite{wang2023review}, and mathematical reasoning \cite{ahn2024large}. 
Beyond performing individual tasks, these capabilities have also enabled the development of LLM-based agents: systems that integrate language models with mechanisms for planning, memory, and interaction with their environment.
LLM agents have demonstrated their ability to solve complex tasks by coordinating, planning, and making decisions \cite{wang2024survey}, and to adapt their behavior in response to ongoing interactions. 
Because of their versatility, LLM agents are increasingly being incorporated into a wide range of applications, including social simulations \cite{rossetti2026ysocial, gurcan2024llm, tornberg2023simulating}.
In social contexts, agents can participate in online discussions \cite{wischnewski2024agree}, operate accounts on real platforms \cite{yang2024characteristics} such as X (formerly Twitter), and LLM populations are employed to simulate complex social dynamics \cite{rossetti2026ysocial}. 
LLM agents are increasingly serving as social actors \cite{heaton2025chatgpt, flint2026group, ashery2025emergent}, interacting with (and affecting) other people or agents. 
One concern with deploying these LLM agents on social platforms is that they can convincingly mimic real individuals and be more persuasive than humans \cite{song2025multi}. 
Moreover, it is still unclear what their collective behavior, capabilities, and intrinsic biases are, and what may emerge at scale \cite{flint2026group}.
\\
LLM populations exhibit human-like collective behaviors such as scale-free network formation \cite{de2023emergence} and information diffusion patterns \cite{chuang2024simulating, park2024generativeagentsimulations1000}. 
They exhibit social conventions and collective preferences~\cite{ashery2025emergent} at the population level, as well as homophily \cite{mehdizadeh2025homophily}, social influence \cite{flint2026group}, and polarization in networked environments~\cite{hashemi2026empirical}.
LLMs agents can reproduce personae with specific demographic and psychological traits, occupations, political leanings, and, more generally, aspects of human behavior. 
When employed to simulate opinion dynamics, LLM agents spontaneously converge to consensus \cite{zomer2026unraveling}. 
However, they tend to overestimate homophily \cite{chang2025llms} and exhibit a bias toward scientific truth \cite{chuang2024simulating}. 
In LLM-human interaction, previous research has assessed that groups of AI agents can exert stronger social influence on users than individual agents, producing larger shifts in users' stated opinions~\cite{song2025multi}.
\\ 
As in human societies, even the collective behavior of LLM depends not only on the agents themselves, but also on the structure of their interactions. 
Prior literature has focused on defining unstructured agent interactions, such as discussions about controversial topics~\cite{taubenfeld2024systematic, breum2024persuasive}.
Recent work has begun to embed LLM agents in more complex social networks and examine how local social influence shapes opinion updating~\cite{mehdizadeh2025your}. 
However, it remains unclear how such opinion dynamics interact with structural properties of the underlying network, particularly homophily~\cite{mcpherson2001birds} and group-size imbalance.
When we interact with others, we form social ties that can be modeled as graph edges, with individuals as nodes. 
Human societies are grounded in unwritten social rules, such as homophily, i.e., the idea that \textit{similarity breeds connection}~\cite{mcpherson2001birds}. 
People naturally tend to gravitate toward those perceived as similar, thereby strengthening their connections with them, while avoiding those perceived as too different.
This behavior has various consequences: it might amplify social divisions, strengthen boundaries between groups, and eventually lead to the segregation of populations into homogeneous clusters.
In LLM societies, the role of such fundamental structural patterns in shaping the evolution of opinion remains unclear.
}
\\\\
\textbf{Why do these questions matter?} There are at least two reasons why the effect of social structure on opinion dynamics of LLM societies needs to be studied. 
By observing how structured populations of LLM agents respond to topological factors such as homophily and class imbalance, one can characterize population-level behavioral tendencies and biases that cannot be inferred from the behavior of isolated agents alone.
This is particularly critical when LLM agents interact in high-stakes contexts \cite{tsvetkova2024new}, such as social networks, and isolating the effects of the models and their topology may allow us to predict the potential outcomes of their interactions with humans.
Second, since LLMs are increasingly employed in agent-based simulations of human social systems~\cite{rossetti2024social, zhang2025socioverse, donkers2025understanding}, it is crucial to evaluate which collective dynamics emerge from the models and which ones are attributable to the structure of the simulated environment.
Recent work has shown that LLM agents can generate homophilic and assortative network structures~\cite{mehdizadeh2025homophily, chang2025llms} and exhibit gender-based homophily \cite{fadaei2026gender}, while other studies have investigated how group size and interaction dynamics can shape collective biases in LLM populations~\cite{flint2026group}.
However, these lines of work have largely considered network structure and population size as separate factors.
\\\\
\textbf{Our contribution.} 
Our work addresses these gaps by introducing simulations in which LLM agents are embedded within a network structure to observe how variations in network homophily and group composition shape their collective outcomes.
{We examine how homophily and relative class sizes within the interaction network affect opinion evolution in LLM societies.
Previous work has investigated the homophilic mechanism in networks generated by LLM agents, focusing particularly on how the attributes of LLM agents (e.g., gender, age, or political leaning) influence link formation, leading to assortative structures and clustered communities~\cite{mehdizadeh2025homophily}.
In contrast, we model populations of LLMs as agents within a preferential attachment framework augmented with homophilic edge formation~\cite{karimi2018homophily}. 
This allows us to manipulate not only traditional network parameters (e.g., degree distribution) but also the social alignment of nodes (via homophily) and the distribution of opinions (via class size imbalance). 
We simulate debates among agents on a given initial statement and trace how opinions evolve over time, with simulations repeated using two different models, which allows structural and model effects to be distinguished.} Our investigation is structured around the following research questions:
\\\\
\noindent\textbf{RQ1.} \textit{How does homophily influence opinion dynamics in balanced LLM agent populations}?
In this experiment, the agent population is evenly divided (50-50) between two opinion classes. We vary the homophily parameter while keeping all other factors constant, particularly class size, to isolate and observe the impact of homophilic attachment on the trajectory and convergence of opinion evolution.
\\\\
\textbf{RQ2.} \textit{How does the effect of homophily on LLM opinion dynamics change under increasingly imbalanced class distributions?}
Building on RQ1, we investigate how skewing the initial class size distribution (i.e., from 70-30 to 90-10) affects the role of homophily in opinion dynamics. This allows us to examine whether the presence of a minority affects the extent to which homophily influences consensus formation and opinion stability.
\\\\
\noindent\textbf{RQ3.} \textit{How does knowledge of neighbors' opinion distribution affect opinion dynamics, at both the collective and the individual level?}
Recent work has shown that LLMs can undergo peer pressure \cite{estrada2013peer} when updating their opinions in response to the opinion distribution of their neighbors \cite{mehdizadeh2025your}. 
We simulate debates between agents with known opinion distributions and assess the impact of this knowledge on opinion change.
\\\\
\noindent\textbf{RQ4. \textit{What micro-level dynamics underlie the aggregate patterns of opinion change?}}
We observe micro-level interactions and model-specific trends by analyzing whether the opinions of interacting agents and the composition of their ego network are associated with statistically significant differences in the probability and direction of opinion change.

{In this study, we hypothesize that by varying network topology while keeping the interaction protocol fixed, we can isolate the contribution of structural factors to collective outcomes. 
We aim to isolate network effects from biases intrinsic to the models themselves.
Network topology strongly affects agents' exposure to opposing opinions: heterogeneous networks facilitate cross-group interactions and can accelerate opinion convergence, whereas higher levels of homophily limit cross-opinion exposure and can preserve more fragmented populations. 
The starting population further modulates these effects, with group-size imbalance producing distinct dynamics depending on the network structure and the LLM employed.
Providing agents with information about their local neighborhood further changes these dynamics, showing that collective behavior depends not only on the underlying network structure but also on the information available to agents during interactions. 
Overall, our results indicate that the collective dynamics of LLM populations cannot be attributed to network structure or model behavior alone, but arise from their interaction.
}
\\ \ \\
The remainder of this work is organized as follows.
In Section \ref{sec:related}, we provide an overview of the literature regarding opinion dynamics and LLM agents. 
In Section \ref{sec:methods}, we introduce the framework for LLM opinion dynamics, and in Section \ref{sec:results}, we introduce the experimental setup and results. 
Finally, in Section \ref{sec:discussion}, we conclude the work by discussing the results and suggesting future research directions.
\begin{figure*}
    \centering
    \includegraphics[width=0.98\linewidth]{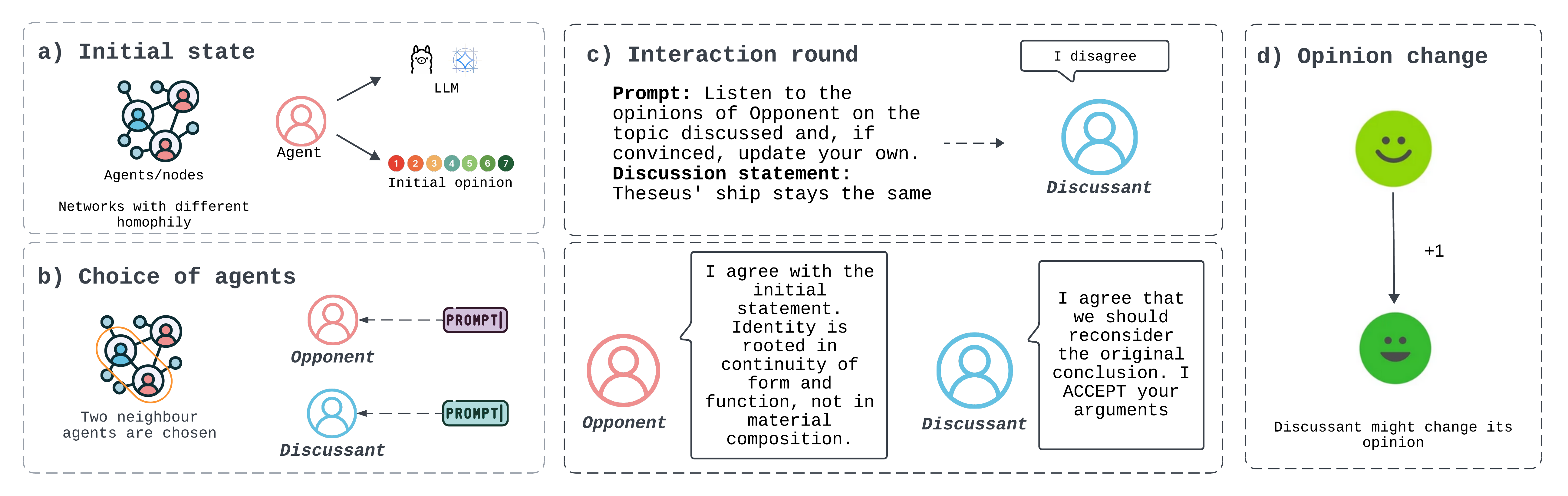}
    \caption{\textbf{Graphical schema of the LLM-OD framework.} The LLM agents population is initialized as nodes in a network; each agent is an LLM instance with an initial opinion in the range [0, 6] (a). At each iteration, two neighbor agents are chosen and prompted to act as \textit{Opponent} and \textit{Discussant} (b). The \textit{Discussant} is prompted to listen to the opinion of the \textit{Opponent} around the discussion statement and may then accept, reject, or ignore such opinion (c) and update their current one accordingly by ±1 (d).}
    \label{fig:frameworkWF}
\end{figure*}

\section{Related Work}\label{sec:related}
\noindent In this section, we discuss relevant literature on Opinion Dynamics and Large Language Model {populations, especially in social simulations}.
\paragraph{Opinion Dynamic Models}
Opinion dynamics (OD) describes computational processes that mimic how individual views evolve over time under empirically observed social influences~\cite{castellano2009statistical,si2018bounded}.
In a typical OD simulation, a network of interacting agents, the \textit{population}, holds opinions encoded as real numbers.
Empirically grounded social mechanisms are cast as update rules, drawing on tools from statistical physics, network science, and probability theory.
For example, an agent may revise its stance by randomly copying a neighbor’s opinion~\cite{holley1975ergodic} (modeling imitation), by averaging the views in its local neighborhood~\cite{deffuant2000mixing} (capturing homophily~\cite{mcpherson2001birds}), or by adopting the most frequent opinion around it~\cite{krapivsky2003dynamics} (reflecting peer pressure and mutual awareness~\cite{crespi2013public}).
OD research asks whether the interplay of these mechanisms drives the system toward one of three mutually exclusive collective states: (i) \emph{consensus}, when all agents share the same view; (ii) \emph{polarization}, when opinions cluster around two opposing poles; or (iii) \emph{fragmentation}, when more than two distinct clusters form~\cite{noorazar2020recent,sirbu2017opinion}.

OD models are also classified by how they encode the opinion variable.
In \textit{discrete} models, opinions can take only a finite set of values, with rules specifying switches among them.
The classic Voter Model~\cite{clifford1973}, for instance, considers populations in which agents adopt one of two opposing stances (e.g., agree/disagree).
By contrast, \textit{continuous} models treat opinions as points on an uninterrupted spectrum.
{In the} DeGroot Model~\cite{deGroot1974reaching}, each individual holds {a subjective probability distribution and each agent updates its opinion as a weighted average of its neighbors' opinions, simulating social influence}.

\paragraph{Large Language Models}
Recent studies have examined the dynamics that arise from interactions among large language models (LLMs). 
These LLM agents typically exhibit realistic behavior both individually and collectively~\cite{park2023generative, park2024generativeagentsimulations1000}, with human-like patterns naturally emerging from their interactions. 
Examples include the formation of scale-free networks~\cite{de2023emergence}, dissemination of information~\cite{gao2023s}, preferential attachment and homophily~\cite{mehdizadeh2025homophily}.
When supplied with demographic backgrounds and personality traits, LLMs can more accurately mimic human behavior~\cite{park2022social,park2023generative,bhandari2025can}, including human biases and political inclinations~\cite{chuang2024simulating}. 
Integrating LLMs with traditional agent-based modeling approaches~\cite{lorig2021agent}, as demonstrated in~\cite{tornberg2023simulating}, has shown how recommender systems can influence the quality of discourse. 
Specifically, agents exposed to popular content tend to exhibit increased toxicity and interparty engagement, mirroring the toxicity levels seen in U.S. Twitter discourse circa 2019.
Further simulations reveal that confirmation bias among agents contributes to social fragmentation~\cite{chuang2024simulating}, aligning with established findings in opinion dynamics.
LLMs are also capable of emulating persuasive communication~\cite{breum2024persuasive,monti2022language}, generating coherent arguments based on psycho-linguistic theories of opinion change, and can be utilized for comprehensive social media simulations~\cite{rossetti2024social}.
Nonetheless, LLMs have a natural inclination toward factual correctness and often resist generating content that contradicts scientific knowledge unless explicitly prompted~\cite{chuang2024simulating}.
{One common problem of LLM agents in social simulation is \textit{sycophancy}, their \textit{excessive agreement with and flattery of the user}~\cite{cheng2026elephant}, with effects of this characteristic varying across models.}
Additionally, LLM agents generally avoid intense conflict~\cite{cisnerosvelarde2024principlesopiniondynamicsmultiagent,tornberg2023simulating} and display sensitivity toward various social and political issues~\cite{Rozado2024political,farina2023chatgpt}.

\section{LLM-powered Opinion Dynamics}
\label{sec:methods}
We present a general framework for describing how populations of LLM agents develop and revise their opinions through structured interactions~\cite{cau2024bots, cau2025selective}.
A graphical overview is given in Figure~\ref{fig:frameworkWF}.
\\
The system is represented by the tuple:
\[
    \mathfrak{F} = \langle \mathcal{G}, \mathcal{S}, \mathcal{O}, f, T, \mathcal{D} \rangle,
\]
where each element captures a specific component of the simulation:

\begin{itemize}
    \item $\mathcal{G} = \langle V, E \rangle$ denotes the interaction network, where $V$ is the set of LLM agents and $E \subseteq V \times V$ is the set of undirected pairs of agents that can interact. The topology may be fully connected (i.e., mean-field) or follow an empirical or synthetic configuration.

    \item $\mathcal{S}$ is the discussion prompt or proposition that defines the initial subject of deliberation. It establishes the semantic dimension along which opinions are assessed.

    \item $\mathcal{O}$ represents the space of admissible opinions. This space can be discrete (e.g., $\mathcal{O} = \{0, 1\}$) or continuous (e.g., $\mathcal{O} = [0, 1]$), encoding positions from complete disagreement to full agreement with the statement $\mathcal{S}$.

    \item $f: \mathcal{O}^n \times \mathbb{R}^m \rightarrow \mathcal{O}$ is the opinion update function. It specifies how an agent revises its opinion as a function of its current position, the outcome of its interaction, and additional contextual variables in $\mathbb{R}^m$.

    \item $T$ is the temporal domain of the simulation, defined as an ordered sequence of discrete time steps $t \in \mathbb{Z}^+$.

    \item $\mathcal{D}(t) = \{o_i(t) \mid i \in V\}$ denotes the opinion state at time $t$, where $o_i(t) \in \mathcal{O}$ is the opinion of agent $i$ at that time.
\end{itemize}

This framework allows the introduction of a wide variety of network topologies, interaction mechanisms, and opinion evolution rules.  
In what follows, we describe a particular instantiation of the model adopted in our experimental setting.
A general overview of its rationale is provided in Algorithm~\ref{algo:LLMOD}.

\begin{figure}[t]
\scriptsize
\caption{Algorithm for the debate simulation}
\label{algo:LLMOD}
\begin{algorithmic}[1]
\Require Graph $\mathcal{G}=\langle V,E\rangle$; statement $\mathcal{S}$;
initial opinions $\mathcal{D}(0)=\{o_i(0)\mid i\in V\}$;
time steps $T_{\max}$; opinion bounds $[o_{\min},o_{\max}]$
\Ensure Opinion-distribution evolution $\mathcal{D}_{\mathrm{all}}$

\State Initialize $\mathcal{D}_{\mathrm{all}}$ as an empty list
\State Append $\mathcal{D}(0)$ to $\mathcal{D}_{\mathrm{all}}$

\For{$t=0$ to $T_{\max}-1$}
    \State $P \gets [\ ]$
    \Comment{All pairs are formed before the LLM calls}

    \For{each agent $i\in V$ in random order}
        \State Select $j$ uniformly at random from $\mathrm{Neighbors}(i)$
        \State Append $(i,j)$ to $P$
        \Comment{$i$ is Discussant; $j$ is Opponent}
    \EndFor

    \State $\widehat{O} \gets [\ ]$

    \For{each $(i,j)\in P$ \textbf{in batch}}
        \State $(\hat{o}_i,\hat{o}_j) \gets
        \mathrm{debate}\bigl(o_i(t),o_j(t),\mathcal{S}\bigr)$
        \Comment{Both opinions are bounded within $[o_{\min},o_{\max}]$}
        \If{$o_i(t)=o_j(t)$}
            \State $(\hat{o}_i,\hat{o}_j) \gets (o_i(t),o_j(t))$
        \EndIf
        \State Append $(i,j,\hat{o}_i,\hat{o}_j)$ to $\widehat{O}$
    \EndFor

    \Comment{Updates are applied sequentially}
    \For{each $(i,j,\hat{o}_i,\hat{o}_j)\in\widehat{O}$}
        \State $o_i(t+1) \gets \hat{o}_i$
        \State $o_j(t+1) \gets \hat{o}_j$
    \EndFor

    \State $\mathcal{D}(t+1)\gets\{o_i(t+1)\mid i\in V\}$
    \State Append $\mathcal{D}(t+1)$ to $\mathcal{D}_{\mathrm{all}}$
\EndFor

\State \Return $\mathcal{D}_{\mathrm{all}}$
\end{algorithmic}
\end{figure}
\noindent In detail, we define the graph topology, the initial statement, the opinion distribution, and the number of iterations.
Each agent is initialized with a discrete opinion on a 7-point Likert scale, such that 0 corresponds with strong disagreement with $S$, 3 corresponds with neutrality, and 6 denotes strong agreement.
At each time step, for each agent $i \in V$, a neighbor $j$ is randomly selected to engage in a debate on $\mathcal{S}$.
As a result of the debate, $i$'s opinion is either \textit{i)} brought closer to $j$'s if they agree, \textit{ii)} brought farther from $j$'s if they disagree, or \textit{iii)} not modified if $j$'s arguments are ignored.
Opinion updates are stored at each iteration and returned at the end.
\\ \ \\
The debate unfolds as shown in Algorithm~\ref{algo:debate}.
The selected agents are first assigned one of two mutually exclusive roles, namely \textit{Discussant} (agent \textit{i}) or \textit{Opponent} (agent \textit{j}).
The \textit{Discussant} is the agent who starts the conversation by expressing and justifying their opinion on $S$.
The \textit{Discussant} is allowed to change its opinion (after engaging with the \textit{Opponent}).
The \textit{Opponent} is tasked with producing a persuasive argument to convince the \textit{Discussant}. 
Both prompts for \textit{Discussant} and \textit{Opponent} are detailed in Appendix \ref{box:discussant-prompt} and \ref{box:opponent-prompt}, respectively.

\begin{figure}[t]
\scriptsize
\caption{LLM debate with VLLM}
\label{algo:debate}
\begin{algorithmic}[1]
\Require Initial opinions $o_i$, $o_j$; statement $\mathcal{S}$;
opinion bounds $[o_{\min},o_{\max}]$
\Ensure Opinion variations $(\delta_i,\delta_j)$ and conversation logs $L_{i,j}$

\State $\delta_i \gets 0$; $\delta_j \gets 0$

\State \textbf{Opponent's turn:}
\State $a_j \gets \texttt{Opponent}$ generates an argument about $\mathcal{S}$

\State \textbf{Discussant's turn:}
\State $a_i \gets \texttt{Discussant}$ replies to $a_j$ with
\texttt{ACCEPT}, \texttt{REJECT}, or another response

\State \textbf{Opponent's reaction:}
\State $r_j \gets \texttt{Opponent}$ reacts to $a_i$ with
\texttt{ACCEPT}, \texttt{REJECT}, or another response

\If{$o_i \neq o_j$}
    \If{$a_i$ contains \texttt{ACCEPT}}
        \State $\delta_i \gets 1$ if $o_j>o_i$, else $-1$
    \ElsIf{$a_i$ contains \texttt{REJECT}}
        \State $\delta_i \gets -1$ if $o_j>o_i$, else $1$

        \If{$r_j$ contains \texttt{ACCEPT}}
            \State $\delta_j \gets -1$ if $o_j>o_i$, else $1$
        \ElsIf{$r_j$ contains \texttt{REJECT}}
            \State $\delta_j \gets 1$ if $o_j>o_i$, else $-1$
        \EndIf
    \EndIf
\EndIf

\State $\delta_i \gets
\mathrm{clip}(o_i+\delta_i,o_{\min},o_{\max})-o_i$
\State $\delta_j \gets
\mathrm{clip}(o_j+\delta_j,o_{\min},o_{\max})-o_j$

\State \Return $(\delta_i,\delta_j,L_{i,j})$
\end{algorithmic}
\end{figure}
{\noindent The debate proceeds in rounds, with each round consisting of one interaction from each agent. 
The total number of rounds is capped by a maximum limit, \texttt{max\_rounds}, to ensure the debate does not continue indefinitely.
Initially, the \textit{Opponent} is prompted to state and support its
opinion on $\mathcal{S}$\footnote{Generated data and code are available at:\href{https://doi.org/10.5281/zenodo.22938433}{10.5281/zenodo.22938433}}.
The \textit{Discussant} then evaluates the \textit{Opponent}'s argument and then declares whether it \texttt{ACCEPTS}, \texttt{REJECTS}, or \texttt{IGNORES} the \textit{Opponent}'s stance.
If the \textit{Discussant} rejects the argument, the \textit{Opponent} can, at the same time, respond to the \textit{Discussant}'s position by declaring whether it accepts, rejects, or ignores it. 
Otherwise, the \textit{Opponent} concludes the exchange with \texttt{END}. 
The \textit{Discussant}'s opinion shift, $o_i^{\Delta+}$, is determined by its response: \texttt{ACCEPT} moves its opinion one step toward the \textit{Opponent}'s opinion, \texttt{REJECT} moves it one step away, and \texttt{IGNORE} leaves it unchanged. 
The \textit{Opponent}'s opinion may also shift in response to the \textit{Discussant}'s reply: acceptance moves it one step toward the \textit{Discussant}'s opinion, rejection moves it one step away, and ignoring the reply leaves it unchanged. 
Opinion shifts are bounded by the minimum and maximum values of the opinion scale.}

\section{Results}\label{sec:results}
We set up multiple simulation scenarios based on the LLM opinion dynamics framework described in Section \ref{sec:methods}.
{For each of the five homophily levels $h \in \{0, 0.25, 0.5, 0.75, 1\}$, we generate a distinct network using the BA-homophily model described in Appendix~\ref{app.networks}. 
The discussion prompt is based on the paradox of Theseus’ Ship, first reported in Plutarch’s \textit{Parallel Lives}~\cite{2018parallel}, which questions whether an object that has had all its components replaced retains its identity. 
In this work, we adopt a positive framing of this paradox (i.e., the ship is the same). 
The prompt used in the simulations is provided in Appendix \ref{app.prompts}.
In a previous study, we examined the impact of a negative framing of the paradox (i.e., "the ship is different") and found that the prompts' effects depended heavily on the model. 
Due to computational constraints, we limited our experiments to small-scale models and a total of 100 agents. 
We performed simulations using two different LLMs: the first set uses Llama 3.2 3B Instruct\footnote{\url{https://huggingface.co/meta-llama/Llama-3.2-3B-Instruct}}, the second, Gemma 3-4B\footnote{\url{https://huggingface.co/google/gemma-3-4b-it}}.
Both models were deployed locally using vLLM \cite{kwon2023efficient} to parallelize the computation.
The temperature of both models was set to $0.8$ to introduce variability in the arguments generated during the debates. 
The maximum generation length was set to 2048 tokens.
No fixed random seed is used, as our objective is to characterize the variability of the collective dynamics across independent realizations.
Each run consists of 100 structured iterations, in which each agent served once as the \textit{Discussant} and interacted with a randomly selected neighbor who served as the \textit{Opponent}. 
\\\\ \noindent\textbf{Experimental design.}
{First, we investigate the effects of different homophily values by considering a symmetric configuration in which both opinion groups are equally represented ($min =  0.5$).
In this scenario, we establish a baseline to evaluate the effects of homophily with a balanced number of agents (50 positive and 50 negative) (\textbf{RQ1}).}
Next, we introduce group imbalance and assess its effect on the population's collective opinion (\textbf{RQ2}). 
We introduce two asymmetric configurations, in which the minority group constitutes either 30\% or 10\% of the population. 
In both cases, the majority is initially assigned the \textit{strongly disagree} opinion, while the minority is assigned \textit{strongly agree}.
This choice is motivated by prior work highlighting LLM agents' tendency to accept positive opinions more readily \cite{cheng2026elephant}. 
By placing the positive opinion in the minority, we can assess whether this directional tendency persists even when the minority is numerically small and whether it can be disentangled from the majority's influence.
Finally, to investigate the role of local social information (\textbf{RQ3}), we repeat the same configurations in a neighborhood-aware condition, in which each \textit{Discussant} receives information about the distribution of opinions in their network during the interaction.
\\
{We characterize the resulting dynamics at both the macro and micro levels. 
At the macro level, we analyze temporal evolution using trajectories showing the mean proportion of agents holding each opinion across 10 independent simulation runs conducted on the same network realization, with a 95\% confidence interval for the mean. 
Thus, the reported variability reflects stochasticity in the opinion dynamics conditional on a fixed network, rather than variability across network realizations.}
At the micro level (\textbf{RQ4}), instead, we estimate conditional transition-probability matrices for the neighborhood-aware condition to examine how the \textit{Discussant}’s and \textit{Opponent}’s opinions, together with the local neighborhood configuration, are associated with opinion shifts.
To assess whether the observed transition probabilities differ from chance expectations, we use a permutation-based null model. 
Specifically, we randomly shuffle post-interaction opinions across interaction events while keeping the pre-interaction opinions of both agents and, where applicable, the neighborhood composition fixed. 
This procedure preserves the marginal distributions and interaction structure while removing systematic associations between the interaction context and the observed opinion changes. 
For each transition-matrix entry, we generate a null distribution from 1.000 permutations and retain only probabilities that differ significantly from the null expectation according to a two-sided test with $p < 0.01$. Further details on the estimation and validation procedure are provided in Appendix~\ref{app.transitions}.

\subsection{RQ1: Effect of homophily on opinion trajectories}
\begin{figure*}
  \centering
  \begin{subfigure}[b]{\textwidth}
    \includegraphics[width=\linewidth]{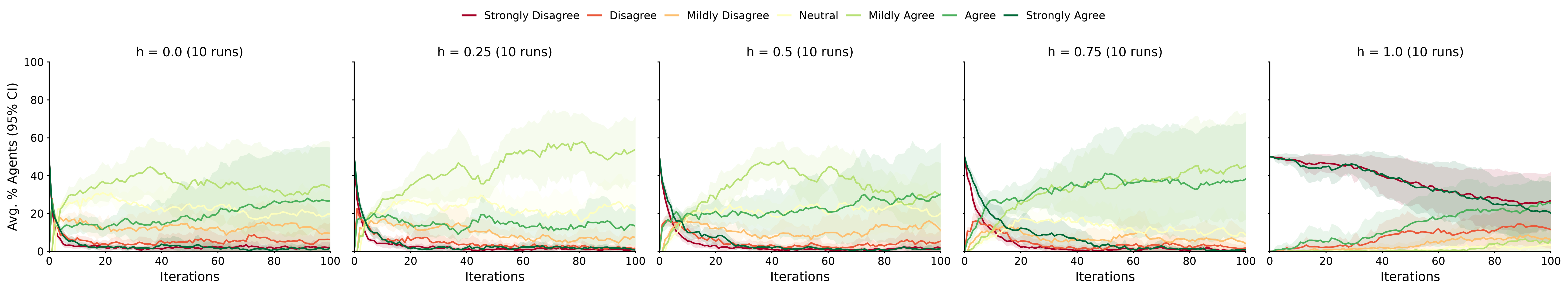}
    \caption{$min{=}0.5$}
    \label{fig:llama05base}
  \end{subfigure}ò\hfill
  \begin{subfigure}[b]{\textwidth}
    \includegraphics[width=\linewidth]{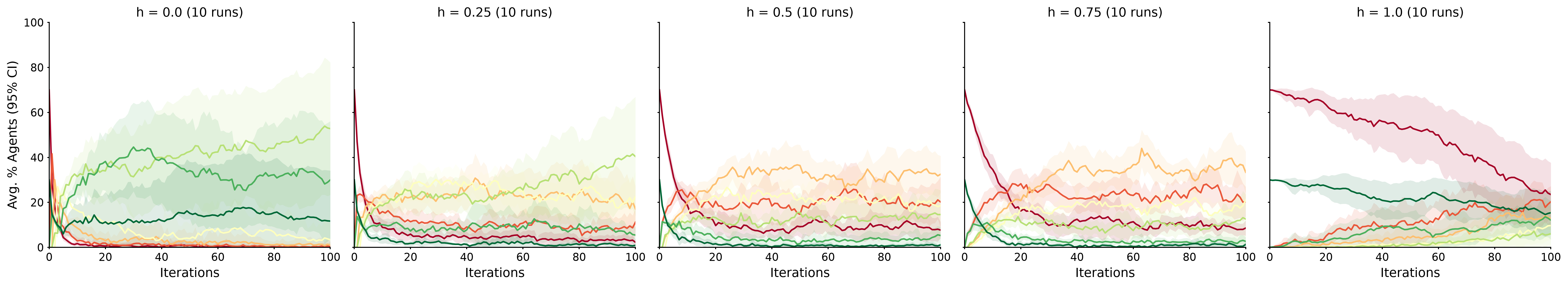}
    \caption{$min{=}0.3$}
    \label{fig:llama03base}
  \end{subfigure}\hfill
  \begin{subfigure}[b]{\textwidth}
    \includegraphics[width=\linewidth]{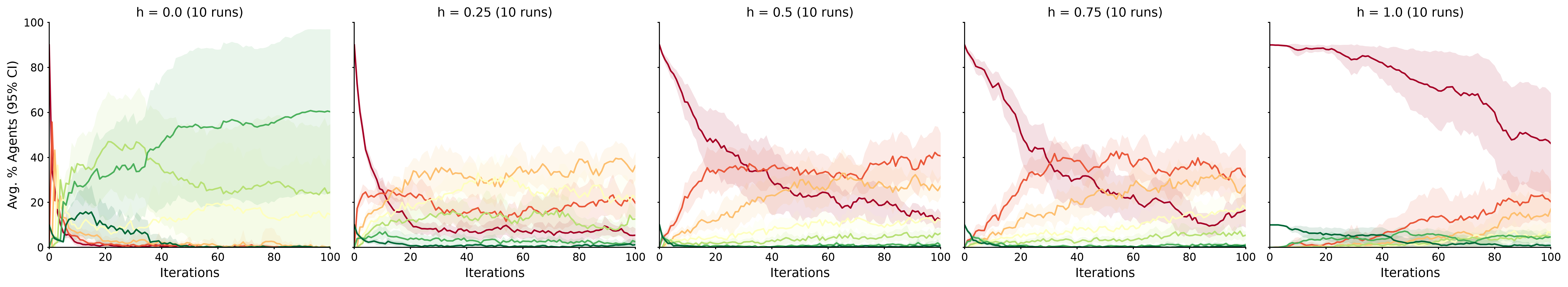}
    \caption{$min{=}0.1$}
    \label{fig:llama01base}
  \end{subfigure}
    \caption{\textbf{Opinion trends for Llama agents across different homophily levels for minority fraction $min = 0.5$ (a), $min = 0.3$ (b), and $min = 0.1$ (c)} 
    Each panel shows the proportion of agents holding different opinion states over 100 iterations.
    All points of the opinion scale are mapped to the colors in the legend, from \textit{Strongly Disagree} (dark red) to \textit{Strongly Agree} (dark green).
}
  \label{fig:min_llama_base}
\end{figure*}

\begin{figure*}
  \centering
  \begin{subfigure}[b]{\textwidth}
    \includegraphics[width=\linewidth]{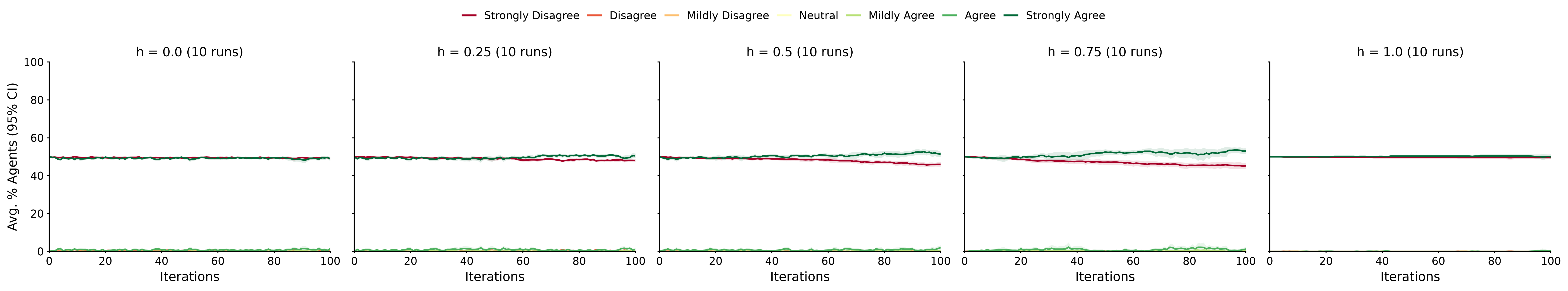}
    \caption{$min{=}0.5$}
    \label{fig:gemma05base}
  \end{subfigure}\hfill
  \begin{subfigure}[b]{\textwidth}
    \includegraphics[width=\linewidth]{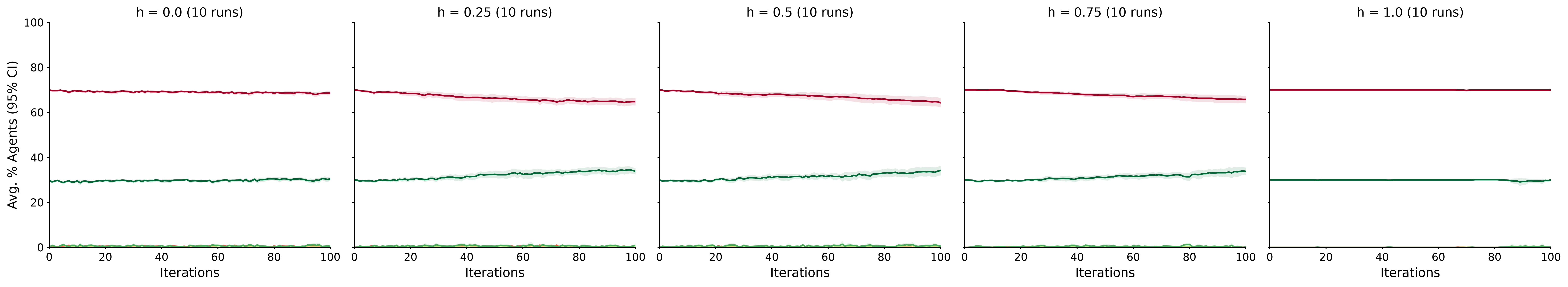}
    \caption{$min{=}0.3$}
    \label{fig:gemma03base}
  \end{subfigure}\hfill
  \begin{subfigure}[b]{\textwidth}
    \includegraphics[width=\linewidth]{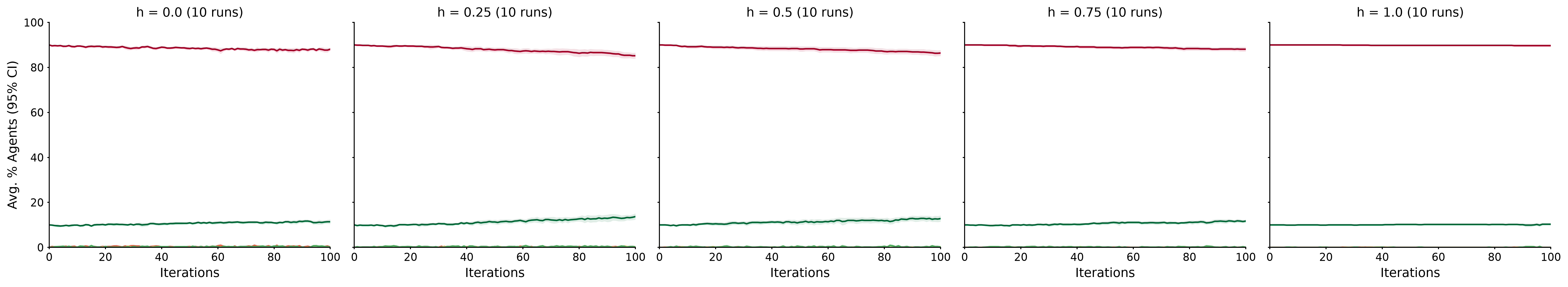}
    \caption{$min{=}0.1$}
    \label{fig:gemma01base}
  \end{subfigure}
    \caption{\textbf{Opinion trends for Gemma agents across different homophily levels for minority fraction $min=0.5 (a)$, $min = 0.3$ (b), and $min = 0.1$ (c)} 
    Each panel shows the proportion of agents holding different opinion states over 100 iterations.
    All points of the opinion scale are mapped to the colors in the legend, from \textit{Strongly Disagree} (dark red) to \textit{Strongly Agree} (dark green).
}
  \label{fig:min_gemma_base}
\end{figure*}
Opinion trajectories for all configurations are shown in Figures~\ref{fig:min_llama_base} and~\ref{fig:min_gemma_base}. 
We first discuss the symmetric population {comparing the two models} (RQ1), and then introduce class imbalance (RQ2).
\\ \ \\
\noindent \textbf{Balanced Scenario (RQ1).}
{The dynamics emerged from Llama and Gemma are markedly different, even in the balanced configuration with equal representation of opinions.
Beginning with Llama (Figure~\ref{fig:llama05base}), all networks with $0 \leq h \leq 0.75$ exhibit a shift from the extreme opinions toward the positive side of the opinion scale.
For $h \leq 0.25$, this shift occurs rapidly during the first 20 iterations, with \textit{Mildly Agree} and \textit{Agree} becoming the most prevalent states.
As homophily increases to $h = 0.5$ and $h = 0.75$, the shift toward positive opinions becomes slower and more variable across runs, although positive intermediate opinions remain prevalent.
Under complete homophily ($h = 1$), the two extreme opinions decline more slowly, and several opinion states coexist at the end of the simulation, resulting in a fragmented population despite agents interacting exclusively with like-minded agents.
}
{In Gemma configurations \ref{fig:min_gemma_base}), the dynamics are unaffected by differences in homophily. 
Opinion trajectories are nearly the same throughout the entire simulation: \textit{strongly disagree} and \textit{strongly agree} each maintain approximately 50\% of the population from initialization to iteration 100, with only negligible fluctuations (Figure~\ref{fig:gemma05base}).}

\subsection{RQ2: Introducing group size imbalance}
{We next examine whether the dynamics observed in the balanced population change when the positive opinion is held by a numerical minority.
For a $70/30$ split, the dynamics remain strongly model-dependent.
For Llama (Figure~\ref{fig:llama03base}), low homophily produces a pronounced shift toward the positive side of the opinion scale.
At $h = 0$, under maximized cross-group contact, \textit{Agree} and \textit{Mildly Agree} agents become the most prevalent, although the population does not reach consensus.
A similar shift occurs at $h = 0.25$, albeit less strongly.
At intermediate homophily levels ($0.5 \leq h \leq 0.75$), the initial \textit{Strongly Disagree} majority decreases substantially, while several moderate and negative opinions coexist.
Under complete homophily ($h = 1$), both initial polarized states are essentially preserved, although both gradually lose agents to intermediate opinion states.
Gemma (Figure \ref{fig:gemma03base}), by contrast, shows stable opinion dynamics, as in the balanced scenario, with variations in homophily producing only negligible changes in the trajectories.
When the minority is further reduced to only 10 agents, the effect of group imbalance becomes more pronounced for Llama (Figure~\ref{fig:llama01base}).
Under complete heterophily ($h = 0$), the population still shifts substantially toward positive opinions, with \textit{Agree} becoming the most prevalent state.
Instead, with $0.25 \leq h \leq 0.75$, the dynamic is more focused on the negative side of the opinion scale.
Under complete homophily ($h = 1$), the initially dominant \textit{Strongly Disagree} group remains prevalent throughout the simulation, while the positive minority remains small.
Gemma exhibits highly stable dynamics (Figure~\ref{fig:gemma01base}).
The initial $90/10$ distribution is almost completely preserved across all homophily levels, with only negligible fluctuations and redistribution toward intermediate positive opinions.
What emerges from both RQ1 and RQ2 is that homophily and group size affect the propagation of opinion change, but the model itself determines whether the shift occurs.
}

\subsection{RQ3: Effect of neighborhood awareness}
Finally, we investigate the effect of neighborhood awareness on each agent. 
To this end, we provide the \textit{Discussant} agent with the distribution of opinions in their ego network for each interaction, so that it can consider changing its opinion based on this{peer-pressure effect. 
For Llama (Figure \ref{fig:llamaopdistr}, neighborhood awareness modifies the dynamics, particularly with cross-group interactions.
In the balanced configuration ($min=0.5$), low homophily reinforces the preference toward the positive side of the opinion scale, even with higher variability across runs.
This trend repeats as homophily increases, reaching $h=0.75$, where the positive cluster remains dominant and more stable across runs.
Under total homophily ($h=1$), the two initial polarized populations coexist, although some move toward intermediate opinions.
Neighborhood awareness affects the distribution of opinions at lower levels of structural segregation.
These effects become even more evident under class imbalance.
With $min=0.3$, low homophily again produces a strong shift toward positive opinions.
At $h=0$, \textit{Agree} becomes the dominant state, reaching approximately 90\% of the population, while the initially dominant \textit{Strongly Disagree} group almost disappears. 
Similar shifts are observed at $h=0.25$, although a substantial minority remains in intermediate states.
Interestingly, for $h=0.5$ and $h=0.75$, however, the population remains fragmented across several opinion states, with \textit{Neutral}, \textit{Mildly Disagree}, and \textit{Disagree} retaining a substantial presence.
Under complete homophily, the initial negative majority remains dominant, although it gradually loses part of its volume to intermediate opinions.
When the positive minority is reduced to $10\%$, the effect of homophily becomes even more evident (Figure~\ref{fig:llama01opdist}).
In all homophily variations $0\leq h \leq 0.75$, the initially small positive group expands rapidly and \textit{Agree} becomes the dominant opinion, while \textit{Strongly Agree} declines.
At $h=1$, the initially dominant polarized groups remain prevalent throughout the simulation, with only a limited shift toward intermediate opinions.
In the case of Llama, local opinion information interacts strongly with network structure. 
Heterogeneous neighborhoods favor large-scale opinion shifts even in the presence of high imbalance, whereas homophily, by increasing the interaction within the same group, limits exposure to opposing groups.
\\
Information about the neighborhood produces a different behavioral response from Gemma (Figure~\ref{fig:gemmaopdistr}).
In the balanced configuration, low and intermediate levels of homophily modulate the trend toward convergence toward positive opinions.
At $h=0.5$, the population has a majority of strongly agreeing agents but retains a substantial negative component.
At $h=0.75$, both extreme opinions remain represented, and under complete homophily ($h=1$), the initial 50/50 split is essentially preserved.
The effect of class imbalance is similarly dependent on homophily.
For $min = 0.3$, low homophily levels ($h=0$ and $h = 0.25$) foster cross-group interaction and produce a redistribution toward positive opinions, with \textit{Strongly Agree} and \textit{Agree} becoming the dominant states.
By contrast, at $h=0.5$ and $h=0.75$, the initially dominant \textit{Strongly Disagree} group remains prevalent, and, under complete homophily, both initial classes remain almost perfectly separated.
For $min=0.1$, this dependence becomes stronger: only under complete heterophily does the small positive minority expand substantially, whereas for $h\geq 0.25$ the negative majority remains dominant and progressively more stable, remaining close to 90\% throughout the simulation.
\\
Overall, introducing this novel information in agents yields the most distinct result from the baseline in RQ1: Gemma, which is fixed across all networks in the baseline condition, finally moves once the neighborhood distribution is present.
This neighborhood effect interacts with homophily and class imbalance to determine whether the minority can propagate through the population.
Under low homophily, both models can exhibit substantial movement toward the positive side of the opinion spectrum, even when the positive stance is initially held by a minority.
However, as homophily increases, cross-group exposure decreases, and the initial majority becomes more stable, particularly in cases of extreme class imbalance.
The main distinction between the models lies in the magnitude and variability of these shifts.
Llama agents tend to redistribute easily across intermediate opinion states, whereas Gemma concentrates the population in a smaller set of dominant opinion states.}

\begin{figure*}[htbp]
  \centering
  \begin{subfigure}[b]{\textwidth}
    \includegraphics[width=\linewidth]{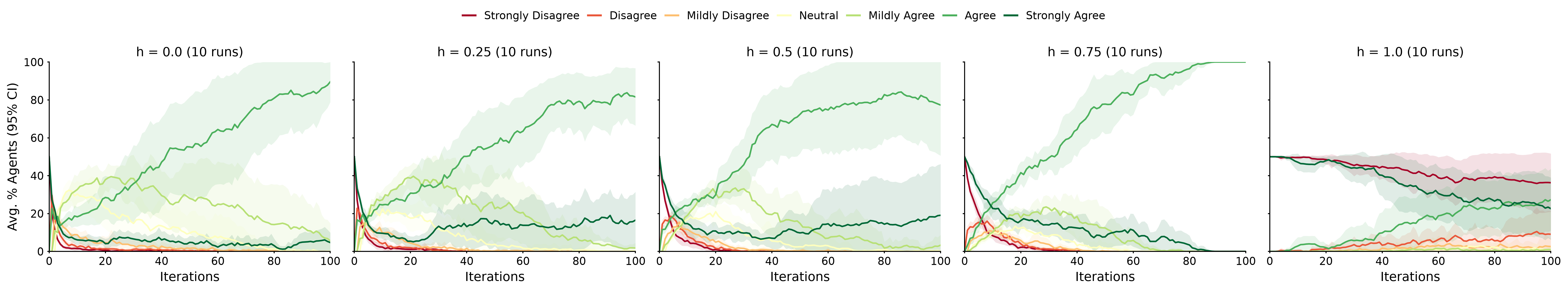}
    \caption{$min{=}0.5$}
    \label{fig:llama05opdist}
  \end{subfigure}\hfill
  \begin{subfigure}[b]{\textwidth}
    \includegraphics[width=\linewidth]{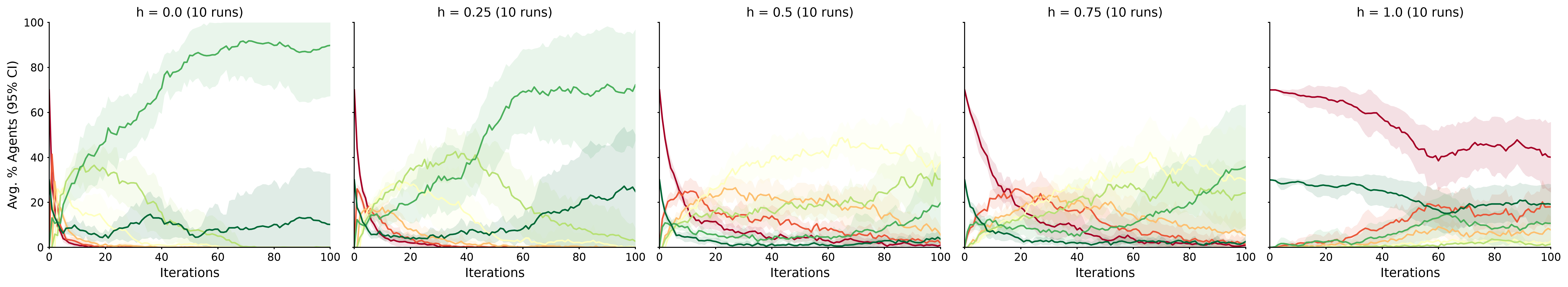}
    \caption{$min{=}0.3$}
    \label{fig:llama03opdist}
  \end{subfigure}\hfill
  \begin{subfigure}[b]{\textwidth}
    \includegraphics[width=\linewidth]{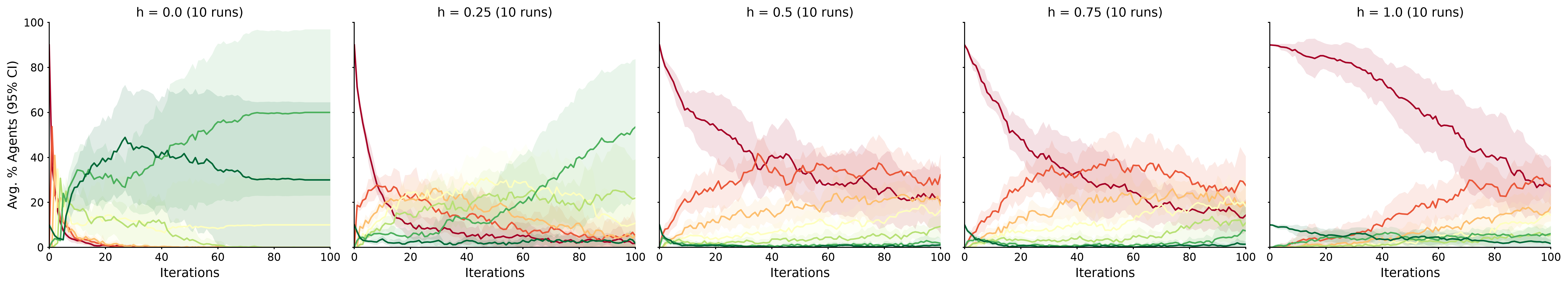}
    \caption{$min{=}0.1$}
    \label{fig:llama01opdist}
  \end{subfigure}
    \caption{\textbf{Opinion trends for Llama agents under neighborhood opinion awareness, across homophily levels, for minority fraction $min=0.5$ (a), $min = 0.3$ (b), and $min = 0.1$ (c)} 
Each panel shows the proportion of agents holding different opinion states over 100 iterations.
All points of the opinion scale are mapped to the colors in the legend, from \textit{Strongly Disagree} (dark red) to \textit{Strongly Agree} (dark green).
}
  \label{fig:llamaopdistr}
\end{figure*}

\begin{figure*}[htbp]
  \centering
  \begin{subfigure}[b]{\textwidth}
    \includegraphics[width=\linewidth]{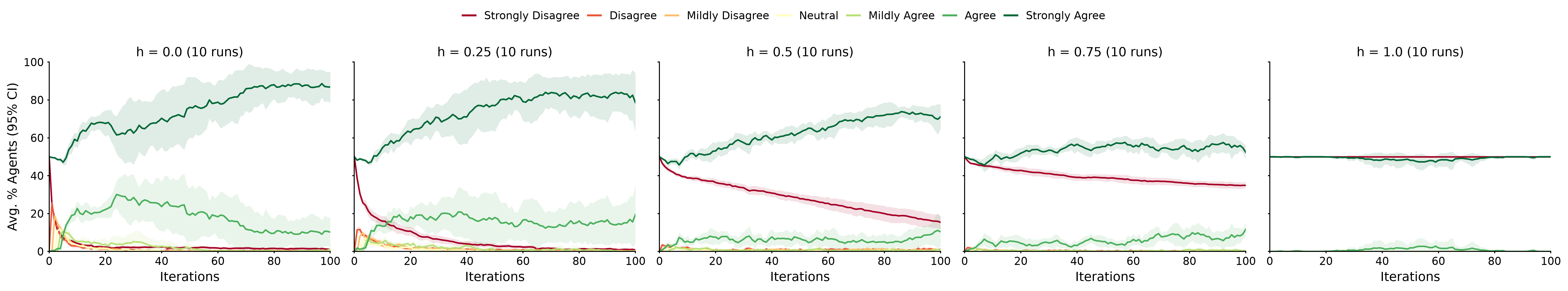}
    \caption{$min{=}0.5$}
    \label{fig:gemma05opdist}
  \end{subfigure}\hfill
  \begin{subfigure}[b]{\textwidth}
    \includegraphics[width=\linewidth]{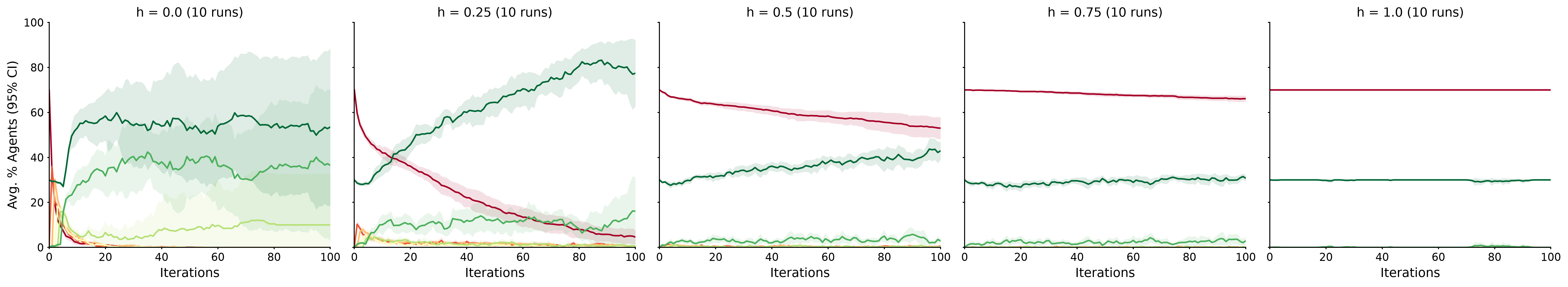}
    \caption{$min{=}0.3$}
    \label{fig:gemma03opdist}
  \end{subfigure}\hfill
  \begin{subfigure}[b]{\textwidth}
    \includegraphics[width=\linewidth]{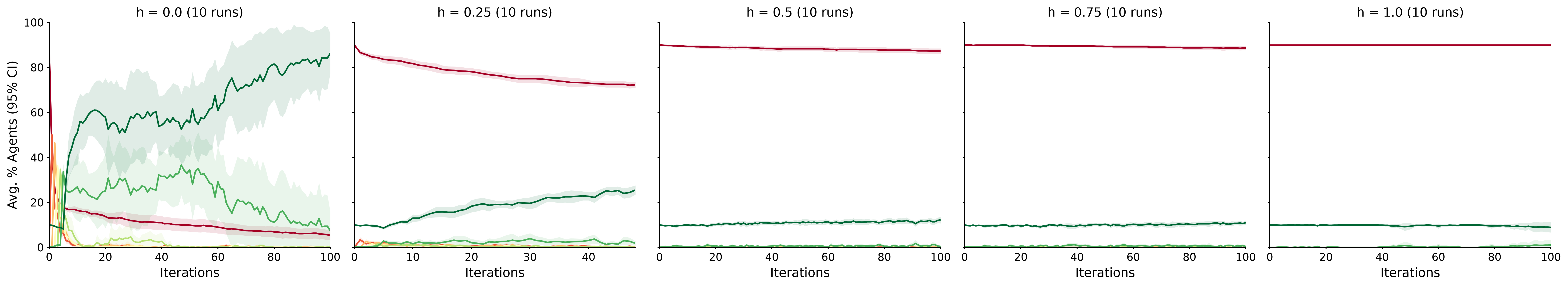}
    \caption{$min{=}0.1$}
    \label{fig:gemma01opdist}
  \end{subfigure}
    \caption{\textbf{Opinion trends for Gemma agents under neighborhood opinion awareness, across homophily levels, for minority fraction $min=0.5$ (a), $min = 0.3$ (b), and $min = 0.1$ (c)} 
    Each panel shows the proportion of agents holding different opinion states over 100 iterations.
    All points of the opinion scale are mapped to the colors in the legend, from \textit{Strongly Disagree} (dark red) to \textit{Strongly Agree} (dark green).
    }
  \label{fig:gemmaopdistr}
\end{figure*}

\subsection{RQ4: Micro-level interaction dynamics}\label{sec:micro}
We first examine whether opinion transitions depend on the initial opinions of the \textit{Discussant} and \textit{Opponent}. 
For each interaction, we estimate the conditional probability that the \textit{Discussant} changes its opinion toward the \textit{Opponent},
\[
P(\text{disc}\rightarrow\text{opp}\mid o_{\text{disc}},o_{\text{opp}}),
\]
where the seven-point opinion scale is collapsed into low ($0$--$3$) and high ($4$--$6$) macrostates. 

{Starting with Llama (Figure \ref{fig:matrices}, first and second row), opinion change is primarily associated with cross-group interactions, while encounters with agents from the same opinion class are less likely to result in a transition between classes.
In the balanced condition (first row), both $low \rightarrow high$ and $high \rightarrow low$ remain high and relatively stable across homophily levels, whereas $low \rightarrow low$ and $high \rightarrow high$ are substantially lower.
The same general pattern is observed under neighborhood awareness, but $low \rightarrow high$ becomes slightly preferred over $high \rightarrow low$, suggesting that information about the local opinion distribution changes how agents respond to opposing opinions, increasing the tendency of agents with low opinions to move towards high-opinion on the scale.
The effect is particularly visible at low and intermediate homophily, whereas at $h=1$ both cross-opinion transitions become less frequent.
Gemma (Figure \ref{fig:matrices}, third and fourth row) shows a different pattern: in the baseline condition, opinion changes are almost nonexistent and reflect the flat trajectories reported above.
Under neighborhood awareness (Figure \ref{fig:matrices}, bottom row), $low \rightarrow high$ transitions under low homophily become slightly more frequent, but as homophily increases, both $low \rightarrow high$ and $high \rightarrow low$ progressively decrease.}
\begin{figure*}
    \centering
    \includegraphics[width=0.9\linewidth]{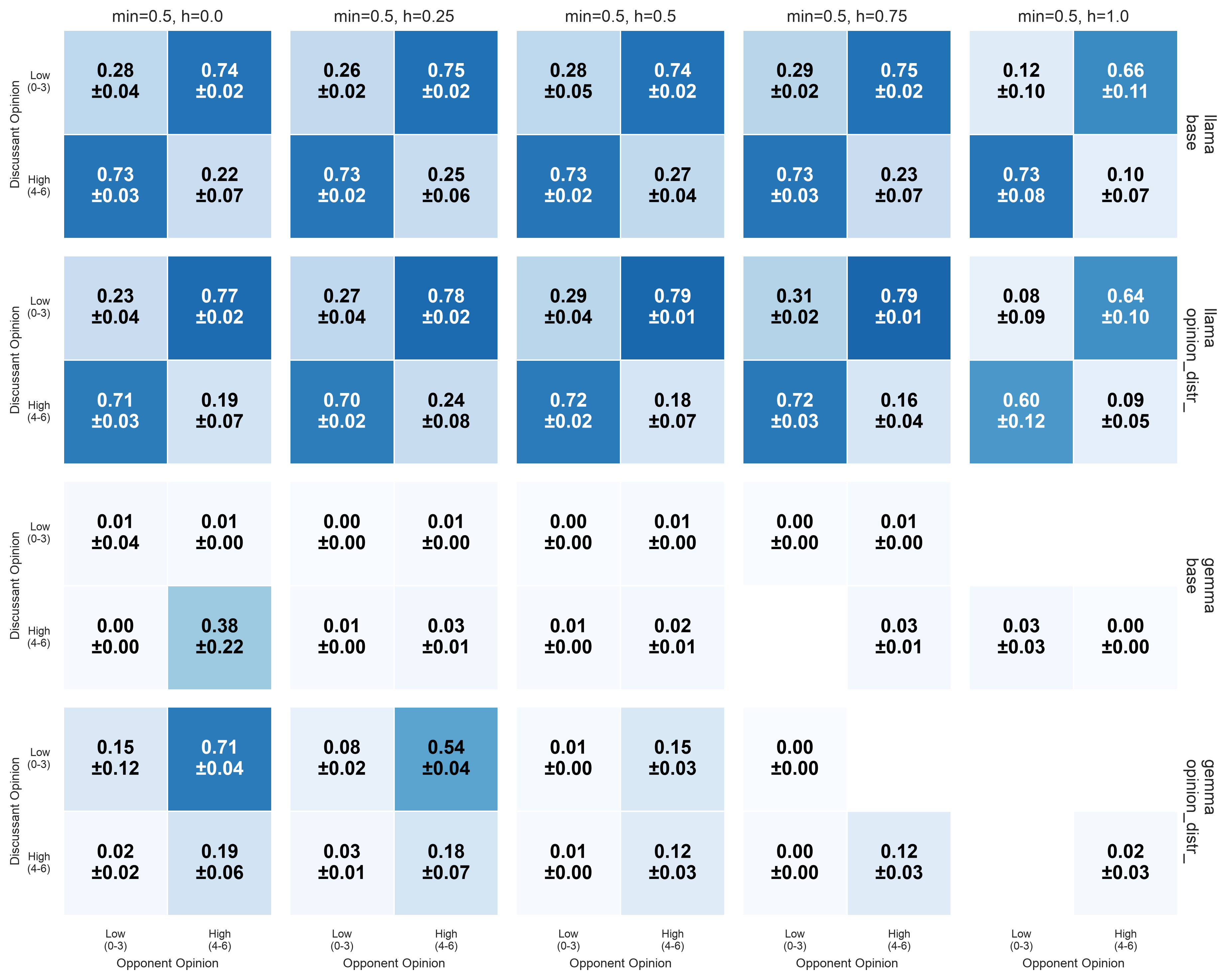}
    \caption{\textbf{Conditional transition probabilities under balanced populations ($\text{min}{=}0.5$) for Llama and Gemma.} Each matrix reports $P(\text{disc}{\rightarrow}\text{opp}\mid o_\text{disc}, o_\text{opp})$, with the 7-point Likert scale collapsed into low- ($0$--$3$) and high-opinion ($4$--$6$) macrostates. Rows correspond to the \textit{Discussant}'s pre-interaction opinion class and columns to the \textit{Opponent}'s. 
    Each matrix represent different homophily level $h\in\{0,\,0.25,\,0.5,\,0.75,\,1\}$;
    Only statistically significant entries ($p{<}0.01$, permutation test) are shown; non-significant cells are left blank.}
    \label{fig:matrices}
\end{figure*}

{To further assess the impact of agents’ knowledge of their neighbors’ opinion distribution, we quantify the probability that the \textit{Discussant} changes its opinion toward the \textit{Opponent}, conditional on the initial opinions of both agents and the prevalent opinion of the \textit{Discussant}’s local neighborhood.
For each condition, we estimate the probabilities for each run; then, the mean across the 10 runs is compared with an aggregated null distribution. 
The matrices in Figure \ref{fig:facet_llama} and \ref{fig:facet_gemma} show the mean and standard deviation between runs for the Llama and Gemma simulations, respectively.

{In this part of the analysis, we consider a \textit{Discussant}'s neighborhood as \textit{supporting} when the majority of its neighbors hold opinions closer to the \textit{Opponent}'s position, \textit{opposed} when the majority hold opinions closer to the \textit{Discussant}'s own initial position, and \textit{mixed} when neighbors are evenly split between the two.
\\\\
For Llama (Figure~\ref{fig:facet_llama}), the probability of changing toward the \textit{Opponent} varies systematically with the configuration of the local neighborhood. 
Variations within the same class are common in opposing and supporting neighborhoods, whereas mixed neighborhoods mitigate these perturbations.
Even if the neighborhood is opposed to the change, cross-opinion transitions remain relatively frequent, particularly for high-opinion \textit{Discussants} facing a low-opinion \textit{Opponent}. 
For low-opinion \textit{Discussants} facing a high-opinion \textit{Opponent}, the probability of changing toward the \textit{Opponent} increases with homophily, reaching its highest value at $h=1$. 
\\
In mixed neighborhoods, the transition probabilities are generally lower and more balanced between the two opinion classes. 
For $h\leq0.75$, both low-to-high and high-to-low transitions are observed, while at $h=1$ statistically significant transitions become more limited.
\\
When the neighborhood supports the potential shift, transition probabilities are substantially higher. 
For low-opinion \textit{Discussants} facing a high-opinion \textit{Opponent}, the probability of changing toward the \textit{Opponent} is close to $1$ across almost all levels of homophily. 
High-opinion \textit{Discussants} facing low-opinion \textit{Opponents} also exhibit high transition probabilities when the neighborhood supports the change, though these probabilities decrease with higher homophily. 
Overall, Llama agents exhibit the highest probability of opinion change when the local neighborhood is aligned with the direction of the potential shift, while changes remain possible, though less consistently, when the neighborhood is mixed or misaligned.
\\\\
For Gemma (Figure~\ref{fig:facet_gemma}), the effect of neighborhood configuration is more pronounced. 
In opposing neighborhoods, cross-opinion transitions are hindered. 
$low \rightarrow high$ transitions are small and become non-significant as homophily increases, whereas $high \rightarrow low$ transitions remain statistically significant at low and intermediate levels of homophily.
At higher homophily, significant transitions become increasingly restricted to configurations involving strong local support. 
Thus, the presence of an opposing neighborhood does not completely prevent a change toward the \textit{Opponent}, but such changes become less prevalent as homophily increases.
\\
Mixed neighborhoods show substantially lower transition probabilities than supporting neighborhoods. 
Most transitions remain small across homophily levels, with only a limited number of statistically significant changes. 
The reduction is particularly evident at high homophily, where significant transitions become rare.
\\
When the neighborhood supports a potential shift in opinion, transition probabilities are generally high. 
In particular, low-opinion \textit{Discussants} facing high-opinion \textit{Opponents} show probabilities close to one across homophily levels, while high-opinion \textit{Discussants} facing low-opinion \textit{Opponents} also exhibit high probabilities of changing toward the \textit{Opponent} when the local neighborhood supports the shift.
\\\\
Overall, Gemma shows a stronger dependence of its dynamics on the local neighborhood configuration than Llama does.
Transitions toward the \textit{Opponent} are most pronounced when the neighborhood is aligned with the direction of the potential shift, whereas opposing and mixed neighborhoods are associated with fewer and weaker transitions.
For Llama agents in opposing neighborhoods, change remains possible but selective: upward transitions from $low \rightarrow high$ persist even when the local environment discourages them, reaching $0.97$ at $h=1$.
Mixed neighborhoods display the most symmetric dynamics, with statistically significant transitions in both directions for $h \leq 0.75$,
In supporting neighborhoods, opinion change largely follows the local consensus: shifts aligned with the neighborhood are reinforced, while the persistence of low-opinion states in $low \rightarrow low$ interactions remains rare. 
Overall, peer pressure acts as a selective amplifier~\cite{moussaid2013social}, strengthening locally consistent changes while still allowing occasional counter-aligned upward moves.}

\begin{figure*}
    \centering
    \includegraphics[width=1\linewidth]{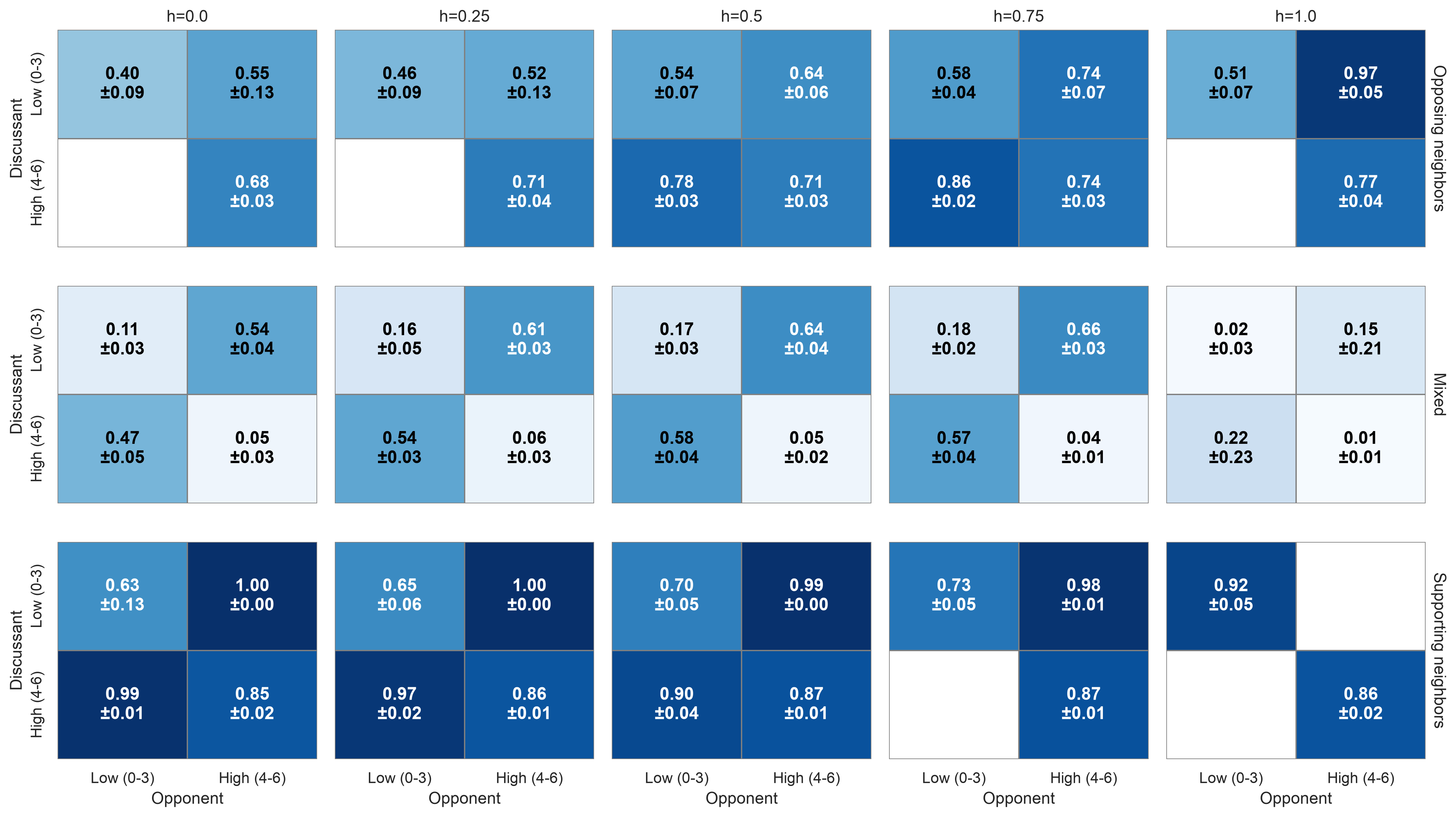}
    \caption{\textbf{Facet plots of statistically significant opinion shifts for Llama agents under different neighborhood conditions. }
    Each panel reports $P\bigl(disc \rightarrow opp \mid o_{disc}, o_{opp}, c\bigr)$, where $c \in \{\text{misaligned}, \text{mixed}, \text{aligned}\}$ denotes the \textit{Discussant}’s local neighborhood relative to the direction of the potential opinion shift.
    Rows correspond to aligned, misaligned, and mixed neighborhoods, while columns vary the homophily level $h$. 
    Only statistically significant transitions are shown.} 
    \label{fig:facet_llama} 
\end{figure*}

\begin{figure*}
    \centering
    \includegraphics[width=1\linewidth]{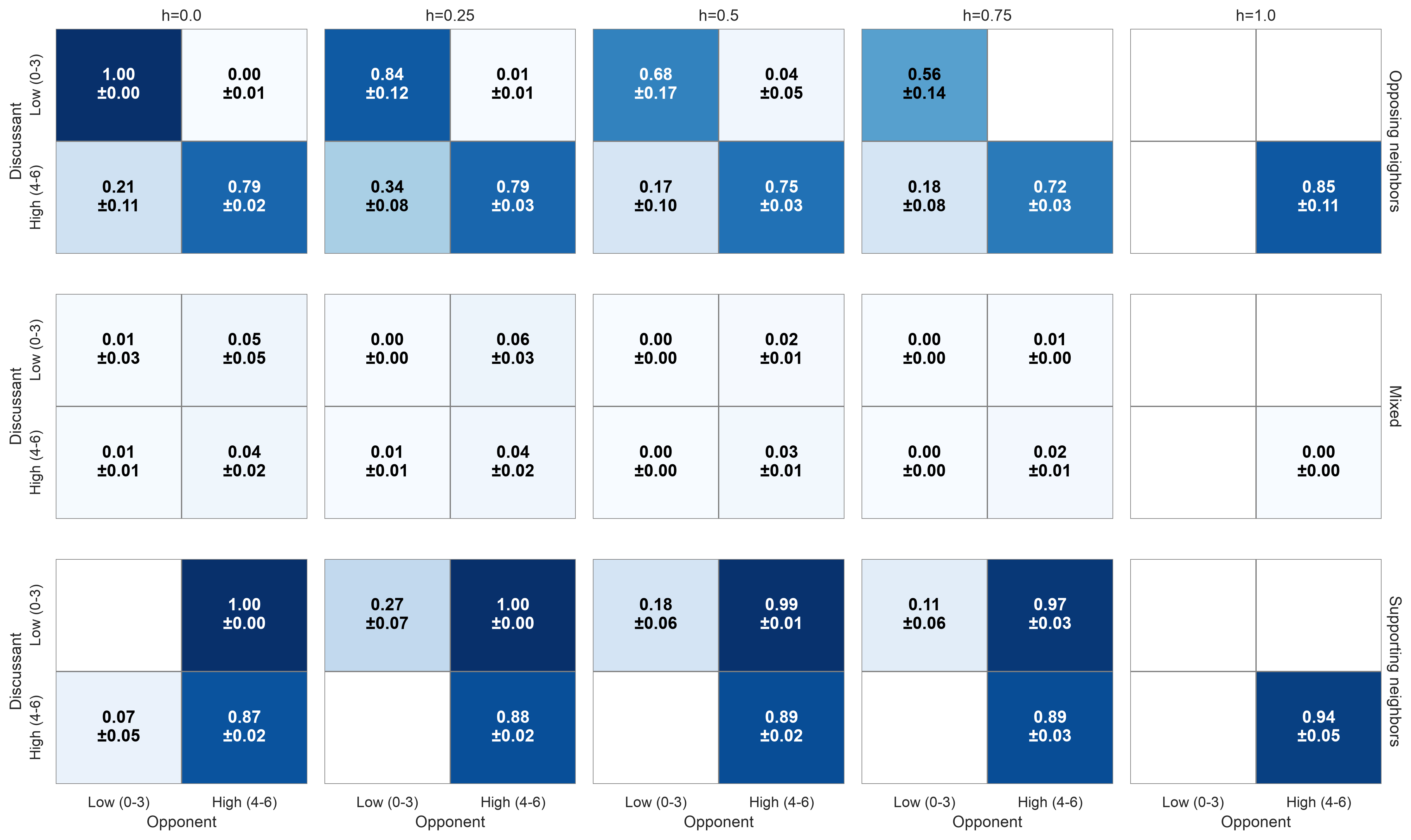}
    \caption{\textbf{Facet plots of statistically significant opinion shifts for Gemma agents under different neighborhood conditions.} Each panel reports $P\bigl(disc \rightarrow opp \mid o_{disc}, o_{opp}, c\bigr)$, where $c \in \{\text{aligned}, \text{misaligned}, \text{mixed}\}$ denotes the \textit{Discussant}’s local neighborhood relative to the direction of the potential opinion shift. 
    Rows correspond to aligned, misaligned, and mixed neighborhoods, while columns vary the homophily level $h$. 
    Only statistically significant transitions are shown.} 
    \label{fig:facet_gemma}
\end{figure*}

\section{Discussion}\label{sec:discussion}
This work investigates how network structure, group composition, and local social information shape opinion dynamics in structured populations of LLM agents.
Overall, our results show that collective outcomes cannot be attributed to network structure alone. 
Instead, they emerge from the interaction between the opportunities for exposure created by the network and the way in which the underlying LLM responds to those interactions. 
This is visible both at the population level, through the evolution of opinion distributions, and at the interaction level, through the conditional probability of opinion change.

\paragraph{Model-dependent collective dynamics.}
A first result is the strong dependence of the dynamics on the underlying language model. Under otherwise comparable experimental conditions, Llama and Gemma produce qualitatively different trajectories. In the baseline condition, Llama agents frequently move away from the initially polarized states, particularly when cross-group interactions are common. Gemma agents, in contrast, largely preserve the initial opinion distribution across homophily levels and group-size configurations. Thus, identical network conditions do not imply comparable collective dynamics across LLM populations.
At the same time, homophily and group imbalance constrain which opinions agents encounter and therefore how far local changes can propagate. Low homophily increases cross-group exposure, whereas higher homophily preserves separation between opinion groups, particularly when one group is numerically small. Network structure and population composition should therefore be interpreted jointly with model-specific responsiveness to social influence.

\paragraph{Local context and micro-level dynamics.}
Neighborhood information further shows that exposure and social context are distinct components of the dynamics. 
Previous work \cite{mehdizadeh2025your} adopted a different perspective, focusing on the intensity of peer pressure necessary for an agent to change its mind, and identifying model-specific conformity thresholds.
However, in our simulations, we examine the direct impact of peer pressure on networks whose homophily and group-size imbalance are explicitly controlled.
Thus, we assess how these topological properties interact with two different models, examining how they shape the models' responses in each scenario. 
Providing agents with the local opinion distribution changes their updating behavior, with substantially stronger effects for Gemma, which is largely static in the baseline condition. 
The transition analysis confirms that local context moderates persuasion: opinion changes are generally more likely when the neighborhood supports the direction of the interaction and less likely when it is mixed or opposed. Importantly, this modulation differs across models. 
Llama remains comparatively responsive across local configurations, whereas Gemma depends more strongly on neighborhood alignment. 
These findings reinforce the broader conclusion that population-level outcomes cannot be attributed to topology alone, but arise from the interaction between structural exposure, contextual information, and model-specific updating behavior.

\paragraph{Limitations and Future Directions.}
Several limitations should be acknowledged. 
First, the experiments rely on a single non-factual discussion topic and on a specific interaction protocol. 
Although this controlled setting facilitates comparisons across experimental conditions, different topics, prompt formulations, or conversational protocols may elicit different updating behaviors. 
In particular, our protocol is asymmetric: the \textit{Discussant} can update after evaluating the \textit{Opponent}'s argument, whereas the \textit{Opponent} can update only after a rejection; the protocol also includes a backfire rule whereby rejection moves an opinion away from the interlocutor. 
We do not test alternative symmetric protocols or a version without backfire, so the extent to which the observed dynamics depend on these design choices remains unknown.
Second, our analysis uses a single network realization for each homophily level. 
Consequently, the reported variability reflects repeated opinion-dynamics runs on fixed network instances and does not quantify variability across alternative networks generated with the same structural parameters. 
Future work should therefore repeat the experiments over multiple independently generated networks for each structural condition.
Third, we consider populations of only 100 agents. This choice limits computational cost but may introduce finite-size effects, particularly in the most imbalanced $90/10$ configuration, where the minority consists of only 10 agents. 
Larger populations may exhibit different patterns of propagation and persistence. 
More generally, alternative network-generating processes, empirical interaction networks, and more heterogeneous topologies may reveal structural effects that are not captured by the BA-homophily model adopted here.
Finally, our comparison is limited to two LLMs. The marked differences between Llama and Gemma show that model choice matters, but they do not establish whether the observed behaviors are characteristic of particular model families, architectures, scales, training procedures, alignment strategies, or prompting conventions. 
Systematic comparisons across a broader range of models, together with variations in topics, interaction protocols, and network realizations, would help distinguish model-specific behavior from patterns that generalize across LLM populations.
}
\appendix
\section{Networks with Controlled Homophily and Group Size}\label{app.networks} To address our research questions, we need to generate networks with controlled group sizes (i.e., node attribute frequencies) and mixing biases (i.e., nodes' preferences for connecting with similar/dissimilar peers).
To do so, we leverage the \textit{BA-homophily} model introduced in~\cite{karimi2018homophily}, which extends the classical Barabási--Albert (BA) preferential attachment model~\cite{karimi2018homophily} by incorporating a \emph{homophily} parameter.
The classic BA model~\cite{barabasi1999emergence} is a foundational generative model for scale-free networks. 
It captures two key mechanisms underlying many real-world complex systems: \emph{network growth} and \emph{preferential attachment}.
The model starts with a small connected network of $m_0$ nodes. 
At each time step, a new node is added to the network and forms $m \leq m_0$ links to existing nodes. 
The probability $\Pi_i$ that the new node connects to an existing node $i$ is proportional to the degree $k_i$ of that node:
\[
\Pi_i = \frac{k_i}{\sum_j k_j}
\]
This rule encodes \emph{preferential attachment}, where well-connected nodes are more likely to attract additional links, a “rich-get-richer” dynamic. 
Over time, this process generates networks with a power-law degree distribution, $P(k) \propto k^{-\gamma}$, with $\gamma \approx 3$.
\\ \ \\
The BA-homophily enriches this framework in two ways. 
First, each node belongs to one of two groups, $a$ (minority) or $b$ (majority), with group sizes defined by relative proportions $f_a$ and $f_b = 1 - f_a$, where $f_a \leq f_b$. 
Second, when a new node $j$ arrives, it selects $m$ existing nodes to connect to, choosing targets based on both their degree and group similarity. 
Formally, the probability $\Pi_{ij}$ that node $j$ connects to node $i$ is given by:

\[
\Pi_{ij} = \frac{h_{ij} \cdot k_i}{\sum_l h_{lj} \cdot k_l},
\]

where $k_i$ is the degree of node $i$ and $h_{ij}$ encodes homophilic preference. 
In the symmetric case where node groups show equal homophilic preferences, we have that
\[
h_{aa} = h_{bb} = h, \quad h_{ab} = h_{ba} = 1 - h
\]
Via this mechanism, the model can generate networks with complete heterophily (nodes only connect to the opposite group, $h = 0$), neutral mixing (no group preference, classical BA, $h = 0.5$), up to complete homophily (nodes only connect within their group, $h = 1$).
As shown in~\cite{karimi2018homophily}, in heterophilic networks ($h < 0.5$), minority nodes become structural hubs, attracting a disproportionate number of links from the majority. Conversely, in homophilic settings ($h > 0.5$), they become increasingly segregated.

\section{Opinion Transition Probabilities and Validation}\label{app.transitions}
To uncover the mechanisms underlying the observed opinion dynamics, we analyze micro-level interaction patterns between \textit{Discussant} and \textit{Opponent} agents. 
Specifically, we construct transition probability matrices that quantify the likelihood that a \textit{Discussant} shifts its opinion toward that of the \textit{Opponent} following an interaction, conditional on the \textit{Discussant}'s and \textit{Opponent}'s initial opinion class. Formally, we estimate:
\[
P\bigl(disc \rightarrow opp \mid o_{disc}, o_{opp}\bigr),
\]
where $disc \rightarrow opp$ denotes a change in the \textit{Discussant}’s opinion toward the \textit{Opponent}’s position, $o_{disc}$ is the \textit{Discussant}’s initial opinion, and $o_{opp}$ is the \textit{Opponent}’s opinion.
By doing so, we aim to estimate the rate of successful (as in persuasive) interactions.
\\ \ \\
For the simulations in which agents are equipped with information on their neighbors' opinion distribution, we additionally condition this probability on the \textit{Discussant}’s local neighborhood configuration, distinguishing whether its neighbors are aligned, misaligned, or mixed, with respect to the direction of change. 
This extension isolates the effect of peer agreement on opinion shifts. Formally:
\[
P\bigl(disc \rightarrow opp \mid o_{disc}, o_{opp}, c\bigr),
\]
where $c \in \{\text{aligned}, \text{misaligned}, \text{mixed}\}$ denotes the configuration of the \textit{Discussant}’s neighborhood relative to the direction of the potential opinion shift,  that is, whether the majority of the \textit{Discussant}’s neighbors hold opinions that are closer to the \textit{Opponent}’s position (aligned), closer to the \textit{Discussant}’s initial position (misaligned), or are evenly split between the two (mixed).
\\ \ \\
To identify statistically significant transitions, we adopt a permutation-based null model that tests whether the observed association between opinion pairs and the direction of change exceeds what would be expected by chance. 
For each interaction, we keep the pair $(o_{disc}, o_{opp})$ (resp. $(o_{disc}, o_{opp}, c)$ for the extension) fixed and randomly shuffle the observed opinion variations across interactions. 
This procedure preserves the empirical distribution of opinion changes and the frequency of each opinion pair, while removing any systematic relationship between them.
For every permutation, we recompute the transition probabilities, thereby obtaining a null distribution for each conditioning configuration. 
The statistical significance of the observed probability is then assessed using a two-tailed test, defined as the fraction of null values that are at least as extreme as the empirical estimate. 

\section{Prompt}\label{app.prompts} 
In this section, we introduce in more detail the prompt given in input to the \textit{Discussant} and the \textit{Opponent} to elicit the debate.
The role of the \textit{Discussant} in the base scenario is modeled as follows: 

\begin{tcolorbox}[colback=gray!5!white,colframe=gray!75!black,title=Discussant Prompt, enhanced, 
    breakable,
    skin first=enhanced,
    skin middle=enhanced,
    skin last=enhanced]
    \label{box:discussant-prompt}
\begin{small}
\begin{verbatim}
[INST] 
    ### You {Discussant_opinion} on the reasoning conclusion 
    provided as input.
    Task:
    - Listen to the argument of {Opponent.name} on the reasoning
    conclusions and decide if you maintain your opinion 
    or change it.
    
    ### Constraints: 
    - At the end of each interaction declare if you 
        - 'ACCEPT' {Opponent.name} argument;
        - 'REJECT' {Opponent.name} argument; 
        - 'IGNORE' your original opinion.
    Write your response with the following format: 
    \"My original opinion was I {Discussant_opinion} 
    on the reasoning. 
    After reading your argument my conclusions are: 
    I <ACCEPT|REJECT|IGNORE> your stance because <argument>\"
[/INST]"
\end{verbatim}
\end{small}
\end{tcolorbox}

The role of the \textit{Opponent} is instead modeled as follows: 
\begin{tcolorbox}[colback=gray!5!white,colframe=gray!75!black,title=Opponent Prompt, enhanced, 
    breakable,
    skin first=enhanced,
    skin middle=enhanced,
    skin last=enhanced]
    \label{box:opponent-prompt}
\begin{small}
\begin{verbatim}
    [INST] 
    You {opponent_opinion} on the reasoning conclusion provided  as input. 
    Support your opinion by providing personal arguments. 
    Avoid using already generated arguments. 
    IF {discussant.name} writes REJECT in his answer, write a second statement 
    where you declare if you <ACCEPT|REJECT|IGNORE> his stance. 
    Otherwise conclude the conversation writing a message with a single word 'END'. 
    \n ### Constraints: \n - In your first statement you must adhere tu your opinion 
    ('{opponent_opinion}')\n - 
    Write your first response with as: \"I {opponent_opinion} on the provided reasoning 
    conclusions. I think that <argument>\" 
    [/INST]",
\end{verbatim}
\end{small}
\end{tcolorbox}

The statement under study is the following:
\begin{tcolorbox}[colback=gray!5!white,colframe=gray!75!black,title= Discussion statement , enhanced, 
    breakable,
    skin first=enhanced,
    skin middle=enhanced,
    skin last=enhanced,]
    Theseus set sail to reclaim the throne as king of Athens. During the journey, parts of Theseus's ship began to break or decay; Theseus and his crew replaced these parts as they sailed. Eventually, each part of the ship is replaced. In the end, the Ship of Theseus is still the same ship on which he originally sailed.
\end{tcolorbox}

Opinion distribution prompts are modeled by giving as input the opinion distribution of the ego network to the \textit{Discussant} as follows:
\begin{tcolorbox}[colback=gray!5!white,colframe=gray!75!black,title= Opinion distribution , enhanced, 
    breakable,
    skin first=enhanced,
    skin middle=enhanced,
    skin last=enhanced,]
\begin{small}
\begin{verbatim}
[INST]
    ### You {discussant_opinion} with the reasoning conclusion provided as input. \n 
    The distribution of opinions in your neighborhood is: {opinion_distribution}. \n 
    Task:\n - Listen to the argument of {opponent.name} on the reasoning conclusion
    and decide whether you maintain your opinion or change it. \n ### Constraints: \n - 
    At the end of each interaction declare whether you: \n   
    - 'ACCEPT' {opponent.name}'s argument;
    - 'REJECT' {opponent.name}'s argument; 
    - 'IGNORE' your original opinion. 
    Write your response in the following format: \"My original opinion was 
    that I {discussant_opinion} with the reasoning. 
    After reading your argument, my conclusion 
    is: I <ACCEPT|REJECT|IGNORE> your stance because <argument>\" 
}
[/INST]
\end{verbatim}
\end{small}
\end{tcolorbox}
% To print the credit authorship contribution details
%% Loading bibliography style file
\bibliographystyle{unsrtnat}

% Loading bibliography database
\bibliography{biblio}

% Biography
%\bio{}
% Here goes the biography details.
%\endbio

%\bio{pic1}
% Here goes the biography details.
%\endbio

\end{document}